\documentclass[twocolumn,trackchanges]{aastex62}
\usepackage{wrapfig,lipsum,booktabs}
\usepackage{natbib}
\usepackage{silence}
\usepackage{amsmath}
\usepackage[flushleft]{threeparttable}
\usepackage[graphicx]{realboxes}
\usepackage{comment}
\usepackage{array}
\usepackage{graphicx}

\usepackage{float}
\usepackage{hyperref}

\shorttitle{Thermodynamic Properties of Early-type Galaxies}
\shortauthors{Frisbie et al.}

\begin{document}

\title{\textbf{Properties of the Hot Ambient Medium of Early-type Galaxies Hosting Powerful Radio Sources}}

\author{Rachel L.S. Frisbie}
\author{Megan Donahue}
\author{G. Mark Voit }
\affil{Physics and Astronomy Department, Michigan State University, East Lansing, MI 48824-2320, USA}
\author{Thomas Connor}
\affil{The Observatories of the Carnegie Institution for Science, 813 Santa Barbara St., Pasadena, CA 91101, USA }
\affil{Jet Propulsion Laboratory, California Institute of Technology, 4800 Oak Grove Drive, Pasadena, CA 91109, USA}
\author{Yuan Li}
\affil{Center for Computational Astrophysics, Flatiron Institute, 162 5th Ave, New York, NY 10010, USA}
\affil{Department of Astronomy, and Theoretical Astrophysics Center, University of California, Berkeley, CA 94720, USA}
\author{Ming Sun}
\affil{ Department of Physics and Astronomy, University of Alabama in Huntsville, Huntsville, AL 35899, USA}
\author{Kiran Lakhchaura}
\affil{MTA-ELTE Astrophysics Research Group, P\'{a}zm\'{a}ny P\'{e}ter s\'{e}t\'{a}ny 1/A, Budapest, 1117, Hungary}
\affil{MTA-E\"{o}tv\"{o}s University Lend\"{u}let Hot Universe Research Group, P\'{a}zm\'{a}ny P\'{e}ter s\'{e}t\'{a}ny 1/A, Budapest, 1117, Hungary}
\author{Norbert Werner}
\affil{MTA-E\"{o}tv\"{o}s University Lend\"{u}let Hot Universe Research Group, P\'{a}zm\'{a}ny P\'{e}ter s\'{e}t\'{a}ny 1/A, Budapest, 1117, Hungary}
\affil{Department of Theoretical Physics and Astrophysics, Faculty of Science, Masaryk University, Kotl\'{a}\v{r}sk\'{a} 2, Brno, 611 37, Czech Republic}
\affil{School of Science, Hiroshima University, 1-3-1 Kagamiyama, Higashi-Hiroshima 739-8526, Japan}
\author{Romana Grossova}
\affil{Department of Theoretical Physics and Astrophysics, Faculty of Science, Masaryk University, Kotl\'{a}\v{r}sk\'{a} 2, Brno, 611 37, Czech Republic}

\begin{abstract}
We present an archival analysis of \textit{Chandra} X-ray observations for twelve nearby early-type galaxies hosting radio sources with radio power $>10^{23} \, \rm{W}~\rm{Hz}^{-1}$ at 1.4 GHz, similar to the radio power of the radio source in NGC 4261. Previously, in a similar analysis of eight nearby X-ray and optically-bright elliptical galaxies, \citet{Werner2012}, found that NGC~4261 exhibited unusually low central gas entropy compared to the full sample. In the central 0.3 kpc of NGC 4261, the ratio of cooling time to freefall time ($t_{\rm{cool}}/t_{\rm ff}$) is less than $10$, indicating that cold clouds may be precipitating out of the hot ambient medium and providing fuel for accretion in the central region. NGC~4261 also hosts the most powerful radio source in the original sample. Because NGC~4261 may represent an important phase during which powerful feedback from a central active galactic nucleus (AGN) is fueled by multiphase condensation in the central kpc, we searched the {\em Chandra} archive for analogs to NGC~4261. We present entropy profiles of those galaxies as well as profiles of $t_{\rm{cool}}/t_{\rm ff}$. We find that one of them, IC~4296, exhibits properties similar to NGC~4261, including the presence of only single phase gas outside of $r \sim 2$~kpc and a similar central velocity dispersion.  We compare the properties of NGC 4261 and IC 4296 to hydrodynamic simulations of AGN feedback fueled by precipitation. Over the course of those simulations, the single phase galaxy has an entropy gradient that remains similar to the entropy profiles inferred from our observations. \\

\end{abstract}

\section{Introduction}

Over the past two decades, \textit{Chandra} has been used to observe the ambient medium of early-type galaxies because of its high sensitivity in the soft X-ray band (0.5-2.0 keV) and its spatial resolution, resulting in 2D spectroscopy of unprecedented quality (e.g. \citealt[][]{KimD2018,DiehlStatler2007,DiehlStatler2008b,DiehlStatler2008a,Lakhchaura2018,Sun2009}). The hot atmospheres of those early-type galaxies have provided key clues about the energetic processes known as ``feedback" \citep[][]{McnamaraNulsen2012,Soker2016,Fabian2012}. 
X-ray signatures of feedback processes observed in the hot atmospheres of nearby, early-type galaxies are also commonly and prominently observed in the hot atmospheres of Brightest Cluster Galaxies (BCG), the brightest and most massive galaxies in galaxy clusters. The supermassive black holes at the center of BCGs in clusters interact with the surrounding medium, inflating bubbles of relativistic plasma (e.g. \citealt[][]{Boehringer1993,Churazov2000,Fabian2003,Fabian2006,Birzan2004,Dunn2006,Dunn2008,Dunn2005,Forman2005,Forman2007,Rafferty2006,McnamaraNulsen2007}). One insight from studying feedback processes in galaxy clusters is that the activity state of the central Active Galactic Nucleus (AGN) in a BCG is closely coupled to the thermodynamic state of the Intracluster Medium (ICM) \citep[e.g.][]{Cavagnolo2008, rafferty2008,VoitDonahue2015,Voit2015Nature}. However, in individual early-type galaxies in groups, like those we discuss in this work, the relationship may be a little more complex \citep[][]{Sun2009,Connor2014}. 
Because the gravitational potential depths are shallower for galaxies in groups than galaxies in clusters, supernova explosions and galactic winds are energetically more important for galaxies than for galaxy clusters. Furthermore, while nearly any reasonable amount of kinetic AGN output can be contained in a cluster atmosphere, the question of whether or not a powerful AGN jet thermalizes its energy output near or far from the AGN depends on the external gas pressure. In turn, the external gas pressure may depend on the large-scale structure the galaxy inhabits.

\citet[][]{McnamaraNulsen2007,McnamaraNulsen2012} have summarized the evidence suggesting that black holes suppress the star formation in massive galaxies, but how the accretion onto the black hole is affected by the surrounding hot gas is less clear. Precipitation-regulated feedback models hypothesize that feedback suspends the ambient medium in a state that is marginally stable to multiphase condensation. Feedback input affects the thermodynamic state and susceptibility of the ambient gas to condensation. Feedback output depends sensitively on the rate at which cold clouds precipitate out of the hot medium \citep[][]{Pizzolato2005,Sharma2012,Gaspari2012,Gaspari2013,Gaspari2015,Gaspari2017,Voit2015b,Voit2017,Wang2018}. Such a system is self-regulated, and finds a balance at the marginally stable point.

Spatially-resolved X-ray spectroscopy of the hot ambient medium provides insight into its thermal evolution. The normalization and shape of an X-ray spectrum yields gas electron density ($n_e$), temperature ($T_{\rm X}$), and metallicity ($Z$). Broadly, for early-type galaxies, the temperature of the hot gas is $\sim$1 keV with a nearly isothermal radial profile, and the radial profile of the electron density approximately follows a power law. The temperature and density of the X-ray gas, considered independently, do not reveal the thermal history because heating and cooling of gravitationally confined gas can cause it to expand or contract without much change in temperature. However, combining these two X-ray observables to make the quantity $K = kT_{\rm X} n_e^{-2/3}$ provides us with more direct information about thermal history, because changes in $kT_{\rm X} n_e^{-2/3}$ correspond directly to changes in the specific entropy of the gas. Only gains and losses of heat energy in the gas can change the entropy, so we can trace the thermal history of the ambient gas of a galaxy cluster by observing the profile $K(r)$, which we will call an entropy profile.

In addition to what we have learned from X-ray observations, numerical simulations show that cool clouds can precipitate out of a galaxy's hot-gas atmosphere via thermal instability even if the galaxy is in a state of global thermal balance, with heating approximately equal to cooling \citep[][]{Mccourt2012,Sharma2012,Gaspari2012}. The critical criterion for precipitation is the ratio between the cooling and free-fall times of the gas. Here, the cooling time ($t_{\rm{cool}}$) is defined to be the time needed for a gas at temperature $T$ to radiate an energy $3kT/2$ per particle, and the free fall time from a galactocentric radius $r$ at the local gravitational acceleration $g$ is defined to be $t_{\rm ff} = (2r/g)^{1/2}$. We note that these models do not presume to claim that the gas must be freely falling. The parameter $t_{\rm ff}$ merely specifies a useful dynamical timescale that characterizes gravitationally-driven motions. The free-fall time does not assume anything about the turbulence, viscosity, or other fluid properties, and is based on galaxy properties that can be inferred from observations of the stellar light.

In both observations and in simulations \citep[][]{Mccourt2012,Sharma2012,Gaspari2012,Li2014a}, cooling appears to be fast enough for a fraction of the hot gas to condense into cold clouds and precipitate out of the hot medium if $t_{\rm{cool}}/t_{\rm ff} \sim$ 10. Precipitation may therefore play an essential role in maintaining the required state of global thermal balance if gas cooled from the hot phase boosts the fuel supply for accretion \citep[][]{Pizzolato2010,Gaspari2013,Gaspari2015,Li2014a,Li2014b}. In numerical simulations, accretion of precipitating clouds can produce a black hole fueling rate two orders of magnitude greater than the Bondi accretion rate of ambient gas. Such strong accretion then produces a feedback response that heats the gas, bringing the system into approximate balance near $t_{\rm{cool}}/t_{\rm ff}\approx 10$. \citet{Voit2015b} showed that early-type galaxies do indeed have $\min(t_{\rm{cool}}/t_{\rm ff}) \approx 10$.  

The hot atmosphere of an early-type galaxy can be broadly categorized as single phase gas or multiphase gas, depending on the extent of the H$\alpha$ and [\ion{N}{2}] emission. Observationally, galaxies with multiphase atmospheres have extended H$\alpha$ and [\ion{N}{2}] emission present outside their centers (central $\sim$kpc), whereas galaxies with single phase atmospheres have no evidence for extended H$\alpha$ emission outside of $\sim \,2$ kpc. X-ray observations of giant ellipticals from \citet[][]{Werner2012,Werner2014} showed that single and multiphase galaxies are distinctly bimodal from $1\textrm{--}10$ kpc. The entropy profiles of single phase galaxies scale as $K\propto r$, while in multiphase galaxies, the entropy scales as $K \propto r^{2/3}$. However, both types exhibit excess entropy in the innermost kpc equivalent to $\sim~\mathrm{2}~\mathrm{keV~cm^{2}}$.

While \citet[][]{Werner2012,Werner2014} showed that both single and multiphase galaxies tend to have entropy excesses relative to a power law in the central kpc, one galaxy differed from the rest. X-ray observations of NGC~4261 from \citet{Werner2012} revealed that the entropy profile of NGC~4261 follows a single power law ($K \propto r$), but instead of exhibiting an excess within the central kpc, the power law continues into the central ${\sim}0.5$ kpc (${\sim}4\arcsec)$. The unusually low entropy in the center ($K\approx0.8\ \mathrm{keV\ cm}^2$) results in $t_{\rm{cool}}/t_{\rm ff}<10$, putting it slightly below the limit at which precipitation appears inevitable. NGC~4261's radio luminosity is 2 orders of magnitude greater than the rest of the \citet{Werner2012} sample, and the central jet power is $10^{44}\mathrm{\ erg\ s^{-1}}$. Adopting a central black hole mass of $M_{BH}=5\times10^8 M_{\odot}$ \citep{Gaspari2013} would require an implausible 30\% mass-energy to jet power conversion efficiency for the radio source to be powered by Bondi accretion alone \citep{Voit2015b}. Simulations from \citet[][]{Gaspari2013,Gaspari2015} showed that a transition to chaotic cold accretion could boost the jet power by up to 100 times over what Bondi accretion of hot ambient gas could achieve and occurs when $t_{\rm{cool}}/t_{\rm ff}\approx10$. 

Because this transitional regime has not been extensively investigated, we decided to explore it by looking for other galaxies like NGC~4261. To that end, we analyzed an archival sample of \textit{Chandra} observations of twelve additional early-type galaxies with powerful radio sources. In this paper, we present a summary of our findings for this archival study, which yielded at least one additional nearby analog, IC 4296, that similarly has both a steep entropy profile with $t_{\rm{cool}}/t_{\rm ff}<10$ at small radii and a powerful radio source. 

The structure of our paper is as follows. Section~2 describes our sample selection, data analysis, and our measurements of the thermodynamic properties. Section~3 presents a comparison of $t_{\rm{cool}}/t_{\rm ff}$ profiles to previous works, a comparison with simulations, and an analysis of the effects of metallicity assumptions on our measurements.  Section~4 concludes by discussing how our sample adds to the paradigm of precipitation-regulated feedback in massive galaxies. Uncertainties are 1-sigma unless otherwise stated, and we assume a $\Lambda$CDM cosmology with $H_0= 70~ \mathrm{km} ~\mathrm{s}^{-1} ~\mathrm{Mpc}^{-1}$ and $\Omega_M=0.3 ~(\Omega_\Lambda= 0.7)$ throughout.

\section{Sample Selection and Data Analysis}

\subsection{Sample Selection and Distances}
\begin{deluxetable*}{lccccccccccc}
\tablecaption{
\label{tab:params}
\textbf{\textit{Chandra} Observations of Early-type Galaxies} }
\setlength{\tabcolsep}{4pt}
\tablehead{ \colhead{Galaxy} & \colhead{$z_{\rm spec}$} & \colhead{$D$} & \colhead{$N_{\rm H,HI}$} & \colhead{1.4 GHz} & \colhead{Obsid} & \colhead{Exp} & \colhead{net cts} & \colhead{Profile} & \colhead{$\alpha$} & \colhead{$\sigma_v$} & \colhead{H$\alpha+$[\ion{N}{2}]}\\
\colhead{}& \colhead{}& \colhead{(Mpc)}& \colhead{($10^{20} \rm{cm}^{-2}$)} & \colhead{($10^{24}\mathrm{\, W \, Hz^{-1}}$)}   & \colhead{} & \colhead{(ks)} & \colhead{per bin} &  \colhead{(Y/N)} & \colhead{(1-10 kpc)} & \colhead{(km $\mathrm{s}^{-1}$)}  & \colhead{morph\tablenotemark{c,}\tablenotemark{d}} } 
\colnumbers \startdata
NGC193 & 0.015 & 63.08 & 2.46 & 0.468 & 11389 & 93.13 & 300 & N & - & 197.6$\pm$4.8 & E\\
NGC315 & 0.016 & 67.20 & 5.88 & 0.973 & 4156 & 53.84 & 930 & N & - & 293.6$\pm$10.1 & U \\
NGC741 & 0.018 & 75.42 & 4.24 & 0.327 & 17198 & 91.02 & 1500 & Y & - & 287.4$\pm$9.3 & N \\
NGC1316 & 0.006 & 25.51 & 1.99 & 9.75 & 2022 & 29.86 & 450 & Y & 0.80$\pm$0.08 & 223.1$\pm$3.3 & E\\
IC1459 & 0.006 & 25.51 & 0.94 & 0.1 & 2196 & 58.00 & 300 & Y & - & 296.1$\pm$6.4 & E \\
NGC3801 & 0.011 & 46.48 & 1.99 & 0.296 & 6843 & 59.20 & 330 & N & - & 191.8$\pm$16.6 & - \\
NGC3894 & 0.011 & 46.48 & 1.83 & 0.125 & 10389 & 38.54 & 300 & N & - & 252.8$\pm$11.3 & -\\
NGC4261 & 0.007 & 31.32 & 1.61 & 2.29 & 9569  & 100.34 & 1600 & Y & 1.09$\pm$0.07 & 296.7$\pm$4.3 & NE \\
IC4296 &0.012 & 50.64 & 3.95 & 5.52 & 3394 & 24.84 & 800 &  Y & 1.12$\pm$0.12 & 327.4$\pm$5.4 & NE\tablenotemark{b}\\
NGC4374 & 0.003 & 18.37 & 2.90 & 0.125 & 803 & 28.46 & 650 & Y & 0.75$\pm$0.05 & 277.6$\pm$2.4 & NE \\
NGC4782 & 0.015 & 63.08 & 3.10 & 3.33 & 3220 & 49.33 & 320 & Y & - & 310.0$\pm$11.3 & NE \\
NGC5419 & 0.014 & 58.94 & 5.40 & 0.146 & 5000 & 14.81 & 320 & Y & - & 344.3$\pm$5.4 & N \\
NGC7626 & 0.011 & 58.34 & 4.59 & 0.222 & 2074 & 26.54 & 370 & Y & - & 266.6$\pm$3.7 & E \\
\enddata
\tablecomments{ Column 1: galaxy name; Column 2: redshift obtained from NED\tablenotemark{a}; Column 3: distance calculated from $z_{spec}$ with the exception of NGC 4374 and NGC 7626, for which we use redshift-independent distances from \citet{Tonry2001_NGC4374} and \citet{Cantiello2007_NGC7626}, respectively; Column 4: galactic neutral hydrogen column densities from \citet{Kalberla2005_HI_column} and \citet{HIP42016_HI_column}; Column 5: radio fluxes from VLA or NVSS \citep{Condon2002} except for NGC 4261 (PKS, \citealt{wright1990PKS}); Column 9 refers to whether there were sufficient counts to make deprojected temperature and density profiles for a galaxy; Column 10:  power-law entropy slope $\alpha$ determined by fitting the relation $K\propto r^{\alpha}$ in the 1--10 kpc interval; Column 11: central velocity dispersion from \citet{Makarov2014_vdisps}; Column 12: H$\alpha+$[\ion{N}{2}] morphology reported by  \citet{Lakhchaura2018} from \citet{connor-inpress} and \citet{sun-inpress}, classified as follows: N: no cool gas emission, NE: H$\alpha+$[\ion{N}{2}] extent $<$ 2 kpc, E: $H\alpha+$[\ion{N}{2}] extent $\geq$ 2 kpc and U: galaxies for which the presence/absence of H$\alpha+$[\ion{N}{2}] could not be confirmed with current observations.}
\tablenotetext{a}{The NASA/IPAC Extragalactic Database (NED) is funded by the National Aeronautics and Space Administration and operated by the California Institute of Technology.}
\tablenotetext{b}{IC 4296 was identified as E in \citet{Lakhchaura2018}. Since its multiphase gas is at $<2$ kpc, we classify it here as NE.}
\tablenotetext{c,}{Based on observations obtained at the Southern Astrophysical Research (SOAR) telescope, which is a joint project of the Minist\'{e}rio da Ci\^{e}ncia, Tecnologia, Inova\c{c}\~{o}es e Comunica\c{c}\~{o}es (MCTIC) do Brasil, the U.S. National Optical Astronomy Observatory (NOAO), the University of North Carolina at Chapel Hill (UNC), and Michigan State University (MSU)}
\tablenotetext{d}{Based on observations obtained with the Apache Point Observatory 3.5-meter telescope, which is owned and operated by the Astrophysical Research Consortium.}
\end{deluxetable*}

NGC~4261 exhibits an unusually low central entropy as well as $t_{\rm{cool}}/t_{\rm ff}<10$ at $r < 0.3$~kpc \citep{Voit2015b}. It also has a powerful radio source emitting $2.3\times10^{24}~\mathrm{W \ Hz^{-1}}$ in the 1.4 GHz band, which may be powered by chaotic cold accretion onto the central supermassive black hole \citep[][]{Gaspari2012,Gaspari2013,Gaspari2015} fed by precipitation of cold clouds out of the hot atmosphere \citep{Voit2015b}. 

In search of other systems similar to NGC~4261, we compiled a {\em Chandra} archival sample of nearby ($z< 0.02$) massive early-type galaxies hosting radio sources with a power output $> 10^{23}\ \mathrm{W\ Hz^{-1}}$ at 1.4 GHz \citep[see Table \ref{tab:params}]{Condon2002}. Other recent studies of \textit{Chandra} observations of early-type galaxies \citep[e.g.][]{Lakhchaura2018,Grossova2019,Juranova2019} include some of the same galaxies, but our sample emphasizes powerful radio sources in order to identify galaxies similar to NGC 4261.  

In our analysis, we used distances derived from the redshifts of the galaxies when calculating the electron density, because the effect of small uncertainties in distance on the inferred density, entropy, and cooling time is small. However, for NGC 4374 and NGC 7626 we used the redshift-independent measurements because the differences between the best redshift-independent distance measurements and the redshift-dependent distances are large ($20-30$\%, see Table \ref{tab:params}).  

In this work, we pay particular attention to NGC~4374 (M84), NGC~1316 (Fornax A), and IC~4296 because \textit{Chandra} observations of the central 10 kpc of those galaxies have the best signal-to-noise ratios among those in our sample, and we are most interested in atmospheric properties closest to the center. Each galaxy represents a different manifestation of a powerful radio source. M84 hosts a Fanaroff-Riley-I radio source\footnote{Abbreviated as FR-I and defined as a radio source in which the low brightness regions of the radio source are further from the galaxy than the high brightness regions \citep{Fanaroff1974}}\citep{Harris2002_M84}. Fornax A has a weak core in the radio (250 mJy), but its radio lobes are some of the brightest radio sources in the sky \citep[125,000 mJy,][]{Ekers1983_FornaxA}. IC~4296 is the Brightest Group Galaxy (BGG) in a nearby galaxy group (Abell 3565), and \textit{Hubble Space Telescope} (\textit{HST}) spectroscopy indicates a central black hole mass of $\sim10^9 M_\odot$ \citep{DallaBonta2009}. Recent \textit{VLA} (Very Large Array) D-configuration observations show improved mapping of the 160 kpc diameter radio lobes, first discovered by \citet{Killeen1986}, located over 230 kpc from the AGN host galaxy as well as \textit{XMM} (X-ray Multi-Mirror Mission) observations that reveal a corresponding X-ray cavity \citep{Grossova2019}.

\subsection{\textit{Chandra} data reduction}
All of the data used in this work are archival \textit{Chandra} data taken between May 2000 and December 2015. All observations were taken with the ACIS-S detector, except for NGC~5419 and NGC~7626, which were obtained with the ACIS-I detector. We reprocessed the archival \textit{Chandra} data listed in Table \ref{tab:params} using CIAO 4.9 and CALDB version 4.7.4. For simplicity, in the case of targets with multiple observations, we chose to analyze the one with the longest net exposure time.

The time intervals containing data with anomalously high background were identified and removed using the  \texttt{deflare} script in CIAO. Bright point sources were identified and removed using the \texttt{wavdetect} script \citep{Freeman2002}. We opted to account for the effect of central point sources in our spatially-resolved spectral analysis. Background images and spectra were derived from the blank-sky fields available from the \textit{Chandra} X-ray Center. The background files contain both particle and photon backgrounds and were filtered and reprojected to match the target observations. We rescaled the reprojected background rates to match the particle count rates, gauged from the event rate between $10.0\textrm{--}12.0$ keV \citep{Hickox2006_Chandrabkg}. Because our analyses are based on regions of the galaxy where the signal is much higher than the background, our results are insensitive to the details of the background scaling.

\subsection{Spectral Analysis}\label{spec_analysis}
We derived deprojected radial profiles of the X-ray gas properties: temperature, density, and gas entropy. To prepare the spectra, we defined radial annuli each containing at least 300 counts after background subtraction (at temperatures around $0.7-1~$keV, a minimum of $\sim300$ counts between $0.5-7$ keV are required for a robust X-ray temperature estimate). We used the definitions of these radial bins to extract radially binned X-ray event spectra for each galaxy and background spectrum from the scaled and reprojected deep background data.

For each galaxy, we fit all radial bins simultaneously with \texttt{XSPEC} v.12.9 \citep{Arnaud1996} using the \texttt{projct} model together with the X-ray thermal emission model \texttt{apec} and Galactic absorption column model \texttt{phabs}. Because the spectral band above $2~$keV is more likely to be dominated by emission from X-ray background and unresolved point sources in typical X-ray spectra of early-type galaxies, we restricted the energy range for the spectral fits to $0.6-2.0~$keV.   

For each galaxy, the Galactic column density and redshift were fixed to the values in Table 1, and the gas metallicity was fixed at a solar abundance. We will discuss the impact of this abundance assumption in Section \ref{metallicity_analysis}. Because the X-ray temperature gradient across the radial range we are interested in is small, we can produce better statistical fits with deprojection by fitting a single temperature across multiple (2-5) adjacent annuli while allowing the spectral normalization to be free in each annulus. The full tabulated results of these fits including uncertainties are provided in Table \ref{tab:radial_profiles}. 

NGC~193, NGC~3801, and NGC~3894 were removed from our sample because there were not enough counts to obtain a deprojected temperature profile with three or more radial bins. NGC~4782 had sufficient counts to extract a profile but had a bright central point source resulting in large uncertainties in the central bins. For NGC~1316 and IC~4296, we do not attempt to fit the central point sources because our primary goal is to assess the shape of the entropy profile and the data quality for future work. Therefore the central $2"$ from IC 4296 and NGC~1316, $0.25~$kpc and $0.12~$kpc, respectively, were excluded from our deprojection analyses of these two galaxies.

\begin{deluxetable*}{lcccccccccc}
\tablecaption{
  \label{tab:radial_profiles}
  \textbf{Radial profile properties for each galaxy with sufficient counts for temperature deprojection.} 
  }
\setlength{\tabcolsep}{4pt}
\tablehead{
\colhead{Galaxy} &  \colhead{radius} &  \colhead{$\Delta r$} &  \colhead{$kT$ bin} &    \colhead{$kT$} & \colhead{$\sigma_{kT}$}  &  \colhead{$n_e$ bin} & \colhead{$n_e$} & \colhead{$\sigma_{n_{e}}$} &  \colhead{$K$} & \colhead{$\sigma_{K}$} \\
\colhead{}& \colhead{(kpc)} & \colhead{(kpc)} & \colhead{ID} & \colhead{(keV)} & \colhead{(keV)} & \colhead{ID} & \colhead{($10^{-2}~\mathrm{cm}^{-3}$)} & \colhead{($10^{-2}~\mathrm{cm}^{-3}$)} & \colhead{($\mathrm{keV~cm}^{2}$)} & \colhead{($\mathrm{keV~cm}^{2}$)} 
} 
\colnumbers \startdata  \\
IC 4296 &   0.48 &  0.24 &          1 &   0.75 &  0.02 &          1 &  16.50 &  0.57 &    2.48 &    0.09 \\
IC 4296 &   0.72 &  0.12 &          1 &   0.75 &  0.02 &          2 &  10.10 &  0.53 &    3.44 &    0.16 \\
IC 4296 &   0.97 &  0.12 &          2 &   0.78 &  0.03 &          3 &   5.62 &  0.39 &    5.29 &    0.33 \\
IC 4296 &   1.45 &  0.24 &          2 &   0.78 &  0.03 &          4 &   3.52 &  0.20 &    7.23 &    0.41 \\
IC 4296 &   1.93 &  0.24 &          3 &   0.84 &  0.03 &          5 &   2.50 &  0.18 &    9.82 &    0.63 \\
IC 4296 &   2.66 &  0.36 &          3 &   0.84 &  0.03 &          6 &   1.19 &  0.06 &   16.08 &    0.86 \\
IC 4296 &   3.87 &  0.60 &          4 &   0.89 &  0.05 &          7 &   0.49 &  0.03 &   30.82 &    1.98 \\
IC 4296 &   6.28 &  1.21 &          4 &   0.89 &  0.05 &          8 &   0.38 &  0.03 &   36.60 &    2.73 \\
IC 4296 &   9.42 &  1.57 &          5 &   2.10 &  1.07 &          9 &   0.24 &  0.03 &  116.60 &   59.92 \\
IC 4296 &  12.56 &  1.57 &          5 &   2.10 &  1.07 &         10 &   0.16 &  0.03 &  152.23 &   79.11 \\
IC 4296 &  15.46 &  1.45 &          6 &   1.29 &  0.21 &         11 &   0.17 &  0.03 &   91.31 &   18.74 \\
IC 4296 &  17.88 &  1.21 &          6 &   1.29 &  0.21 &         12 &   0.17 &  0.05 &   91.62 &   22.83 \\
 &  &  & \nodata  & & \nodata &  &  \nodata & & \nodata &    \\
\enddata
\tablecomments{
    Table~\ref{tab:radial_profiles} is published in its entirety in the machine-readable format. A portion is shown here for guidance regarding its form and content. Errors given for radius represent bin widths, all other errors are 1 sigma. Column 1: galaxy name; Column 2: radial bin center; Column 3: half-width of the radial bin; Column 4: grouping of temperature bins; Columns 5-6: best fit temperatures and their errors; Column 7: electron density bin number; Columns 8-9: best fit densities and their errors; in units of $10^{-2}~\mathrm{cm}^{-3}$ for compactness; Columns 10-11: calculated entropies and their errors.}
\end{deluxetable*}
\subsection{Thermodynamic Properties}
\subsubsection{Electron Density Profiles}
To estimate the electron density within a given concentric shell $i$, we use the best-fit spectral normalization from the deprojection model in \texttt{XSPEC}, 
\begin{equation}
\eta_i=\frac{10^{-14}}{4\pi D^2(1+z)^2}\int n_{e,i} n_{p,i} dV_i
    \; \; .
\label{eq:norm}
\end{equation}

The \texttt{projct} model performs the projection from 3D to 2D, and the total emission measure within the extraction volume as shown in Equation \ref{eq:norm}, in which $D$ is the angular diameter distance to the galaxy in centimeters (Table 1), $n_e$ and $n_p$ are the electron and proton number densities, respectively, in $\mathrm{cm}^{-3}$, and $V_i$ is the volume of the concentric shell in $\mathrm{cm}^{3}$. With this definition of normalization, the expression
\begin{equation}
n_e({\rm shell})=\sqrt{\frac{4\pi\eta({\rm shell})D^2 (1+z)^2}{10^{-14}(n_e/n_p)V({\rm shell})}}
\label{eq:edensity}
\end{equation}
gives us the deprojected radial electron density profile for each galaxy. 

\subsubsection{Entropy and $t_{\rm{cool}}/t_{\rm ff}$ Profiles}
\begin{figure*}
\centering
    \includegraphics[clip,trim=3cm 0cm 1.5cm 0cm,width=\textwidth]{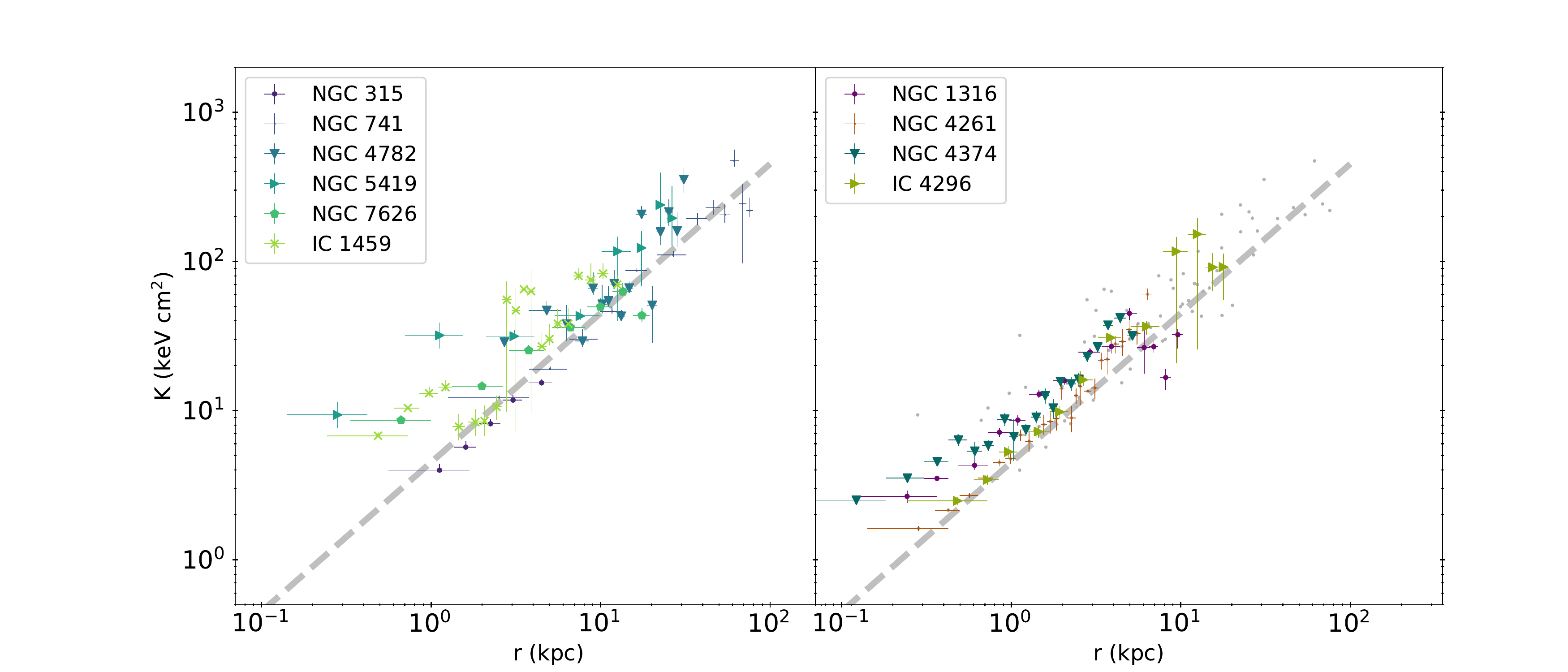}
  \caption{Left panel: Entropy profiles for the galaxies in our sample with sufficient data counts to extract a deprojected radial profile but insufficient data to isolate the central $\sim0.5$ kpc. Right panel: Deprojected entropy profiles of the four galaxies with the best data quality (NGC~4261, IC~4296, NGC~1316, NGC~4374). For comparison, the gray dots are the data points from the galaxies in the left panel. Gray dashed lines on both plots show power-law profiles with $K \propto r$ to illustrate that NGC 4261 and IC 4296 differ from the other galaxies with comparable data quality (NGC 1316 and NGC 4374) by approximately following a similar power law into the central kpc, rather than exhibiting a small excess like the other single phase galaxies.}
  \label{fig:K_profile_all}
\end{figure*}

\begin{equation}
    t_{\rm{cool}}=\frac{3}{2}\frac{nkT}{n_{e}n_{H}\Lambda(T,Z)}
    \label{eq:tctff}
\end{equation}
where $n$ is the total number density of particles, $n_e$ is the electron density, $n_p$ is the hydrogen density (where we assume $n_p=n_e/1.2$), and $\Lambda(T,Z)$ is the temperature dependent cooling function for plasma of metallicity $Z$. Our fiducial cooling function, from \citet{Schure2009}, assumes a solar-metallicity ($Z_\odot$) plasma. The free-fall time is calculated assuming a singular isothermal sphere with velocity dispersions found in Table \ref{tab:params}. We calculated $t_{\rm{cool}}/t_{\rm ff}$ for NGC~4261 and three additional galaxies with the best data quality (NGC~1316, NGC~4374, and IC~4296).

\begin{figure}
\centering
  \includegraphics[clip,trim=1.5cm 2.5cm 2cm 2.5cm,width=0.5\textwidth]{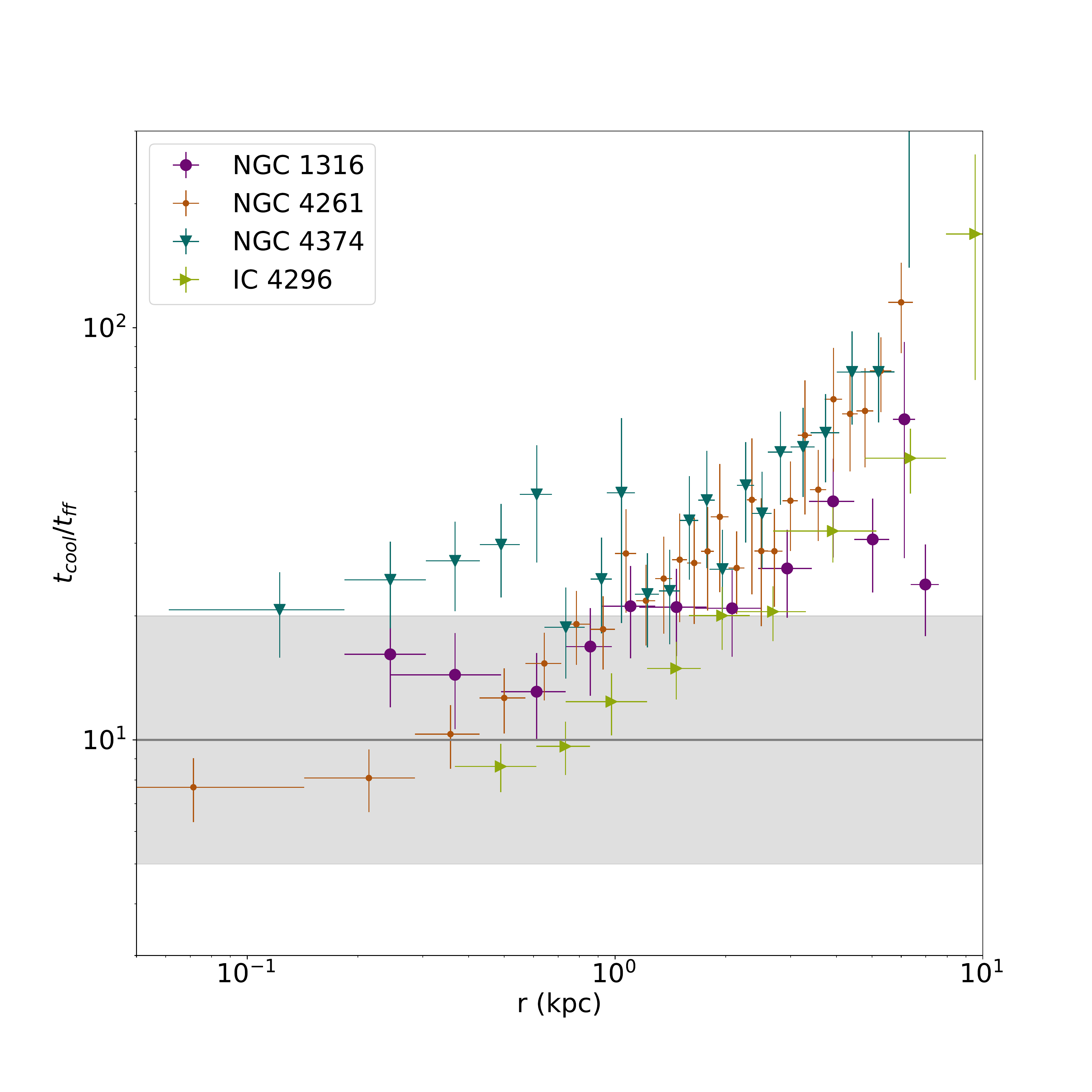}
  \caption{Radial profiles of $t_{\rm{cool}}/t_{\rm ff}$ for the four galaxies with the best S/N. The shaded region ($t_{\rm{cool}}/t_{\rm ff}=5-20$) represents the precipitation zone where multiphase gas is found for $r=1-10\mathrm{kpc}$. We find that, like NGC~4261, IC~4296 reaches $t_{\rm{cool}}/t_{\rm ff}<10$ in the central $\sim1$ kpc while the other galaxies do not.}
  \label{fig:tctff_profile}
\end{figure}

\section{Discussion}
\subsection{$t_{cool}/t_{\rm ff}$ Profiles and Multiphase Gas}
Figure \ref{fig:tctff_profile} shows the $t_{\rm{cool}}/t_{\rm ff}$ profiles for the four galaxies with entropy profiles that come closest to probing the inner $\sim0.5$ kpc of the galaxy. The profiles of IC~4296 and NGC~4374 are of particular interest. While the data are not of the resolution of NGC~4261 they still allow us to see the shape of the $t_{\rm{cool}}/t_{\rm ff}$ and entropy profiles near the central $\sim0.5$ kpc of the galaxy. We also note that while the X-ray structure of NGC 315 is not resolved inside $\sim$1 kpc, its gas entropy profile appears to follow a single power law like IC4296 and NGC 4261. Furthermore, from the spectra reported in \citet[][]{Ho1997,Ho1993}, its multiphase gas appears to be confined to the nucleus, making it another promising candidate for a system in this powerful but possibly short-lived state.

\citet{Voit2015b} showed that $t_{\rm{cool}}/t_{\rm ff}$ in the central $\sim1$ kpc of both single phase and multiphase galaxies usually remains above the apparent precipitation limit at $t_{\rm{cool}}/t_{\rm ff}\sim10$. Further out from the center (1--10 kpc), galaxies with multiphase gas have $t_{\rm{cool}}/t_{\rm ff}$ profiles that approximately track this precipitation limit, whereas galaxies with single phase gas generally lie above the precipitation zone at $t_{\rm{cool}}/t_{\rm ff}\sim5-20$ (blue shaded region in Figure \ref{fig:tctff_profile}). \citet{Voit2015b} found that, in a sample of morphologically relaxed, X-ray bright galaxies \citep{Werner2012}, only the radial profile for NGC~4261 dipped below $t_{\rm{cool}}/t_{\rm ff}\sim10$ in the center. 

In our sample, the $t_{\rm{cool}}/t_{\rm ff}$ profile for NGC 4374 remains above the precipitation zone, and NGC 1316 is consistent with the multiphase galaxy pattern from \citet{Voit2015b}. However, IC 4296 goes down to $t_{\rm{cool}}/t_{\rm ff}\sim10$ near the center, as in NGC 4261, suggesting that the AGN feedback occurring in IC 4296 has interesting similarities to that of NGC 4261. The data were sufficient to probe the inner $\sim0.5$ kpc of NGC~4261, but in general, the profiles more closely follow a single power law than a power law with an excess inside the central kpc.

The H$\alpha$ emission in NGC 4261 is nuclear rather than extended \citep[][]{Ferrarese1996,Lakhchaura2018}, consistent with the picture of giant galaxies with single phase gas having  entropy profiles that scale as $K(r)\propto r$. Of our studied galaxies, IC~4296 most closely resembles NGC~4261, and \citet{Grossova2019} reported that in narrow band images from the \textit{Hubble} and SOAR telescopes, IC~4296 also has no H$\alpha$ emission beyond $r \sim2$ kpc.

\subsubsection{Comparison with Previous X-ray Analysis}

In an independent analysis, \citet{Lakhchaura2018} report entropy profiles for a sample of 49 elliptical galaxies, including eight of the galaxies analyzed in this paper: NGC~315, NGC~741, NGC~1316, NGC~4261, NGC~4374, NGC~4782, IC~4296, and NGC~5419. While there are small variations among bin sizes and radial ranges, we verified that our results are nevertheless mutually consistent within the measurement uncertainties. However, the work of \citet{Lakhchaura2018} treated gas metallicity differently, which we address in Section \ref{metallicity_analysis}. In Section \ref{tctff_analysis}, we include some of the results of \citet{Lakhchaura2018} in our discussion. Additionally, in Table \ref{tab:params}, we report the multiphase gas classifications from \citet{Lakhchaura2018} as well as additional results from \citet{connor-inpress} and \citet{sun-inpress} that use observations carried out using the SOAR optical Imager (SOI) and Goodman High Throughput Spectrograph of the 4.1m SOuthern Astrophysical Research (SOAR) telescope and the Apache Point Observatory (APO) Astrophysics Research Consortium  (ARC) 3.5m telescope.

\subsection{Radio Luminosity and $t_{\rm{cool}}/t_{\rm ff}$}\label{tctff_analysis}
\begin{figure*}
    \centering
    \includegraphics[clip,trim=1cm 2cm 2cm 2cm,width=0.6\textwidth]{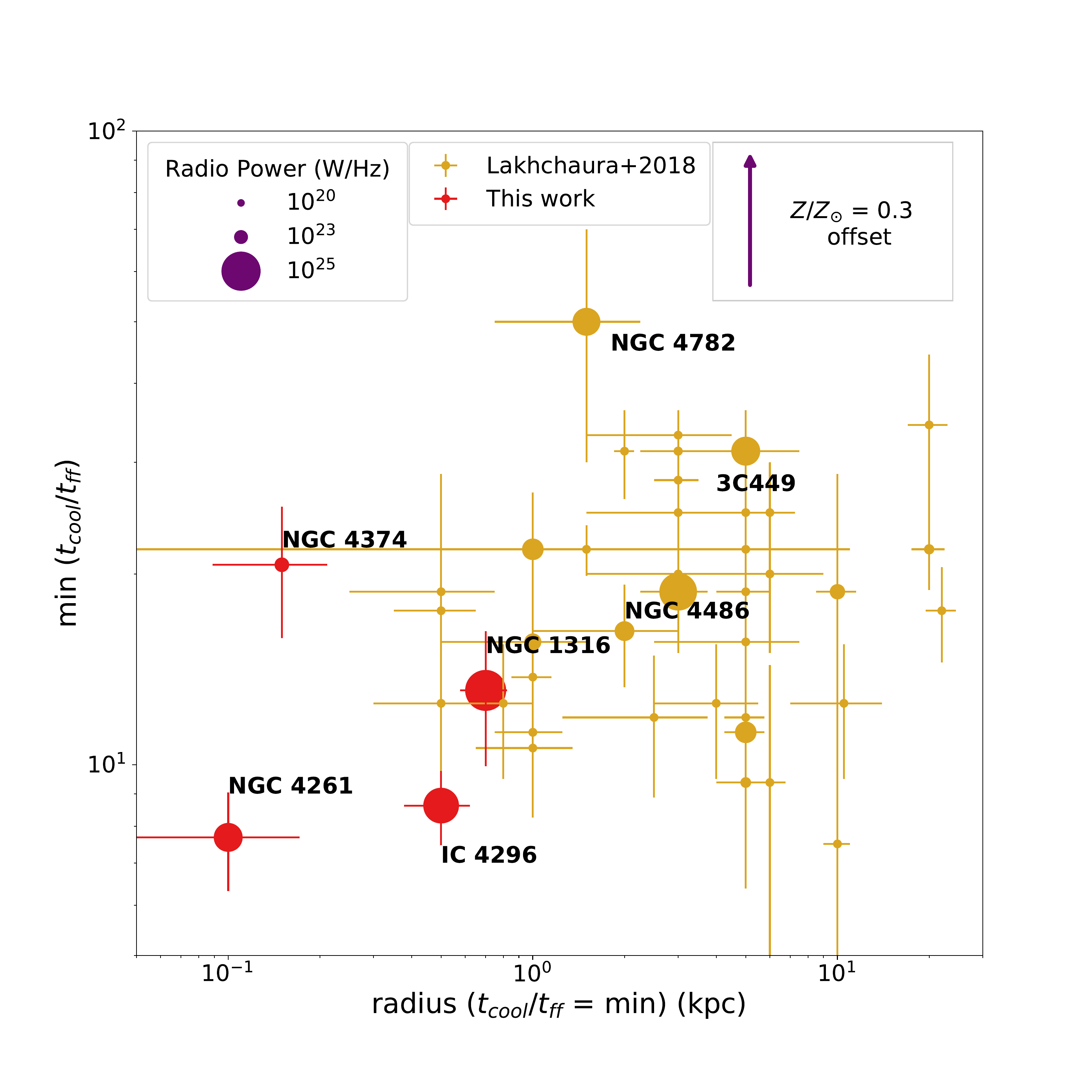}
    \caption{The radius where we measured the minimum value of the $t_{\rm{cool}}/t_{\rm ff}$ profile is plotted against the minimum $t_{\rm{cool}}/t_{\rm ff}$ for the sample of \citet[gold]{Lakhchaura2018} offset to solar metallicity (see \ref{metallicity_analysis}), with our four galaxies of best data quality (red). The purple arrow represents the offset between the adjusted values and the $0.3~Z_{\odot}$ values from \citet{Lakhchaura2018}. The relative size of the points represents their radio power (in $\rm{W}~\rm{Hz}^{-1}$) in the 1.4 GHz band. NGC~4261, IC~4296, and NGC~1316 have a small $\mathrm{min}(t_{\rm{cool}}/t_{\rm ff})$ radius, a low $\mathrm{min}(t_{\rm{cool}}/t_{\rm ff})$, and a greater radio power than most galaxies in the sample.
    } 
    \label{fig:LR_mintctff}
\end{figure*}

Figure \ref{fig:LR_mintctff} shows the minimum values of $t_{\rm{cool}}/t_{\rm{ff}}$ for NGC~4261, IC~4296, NGC~1316, and NGC~4374, along with the giant ellipticals from \citet{Lakhchaura2018}, plotted as a function of the radius at which $t_{\rm{cool}}/t_{\rm{ff}}$ reaches its minimum value. We have adjusted the min($t_{\rm{cool}}/t_{\rm ff}$) values reported by \citet{Lakhchaura2018} for uniform comparison with our work, using the correction factor estimated in Section \ref{metallicity_analysis}. The typical amplitude and direction of that correction is plotted on Figure~\ref{fig:LR_mintctff} in the form of a purple arrow. This adjustment typically decreased the $t_c/t_{\rm{ff}}$ estimates from \citet{Lakhchaura2018} by a factor of 1.6. Points vary in size according to radio power in the 1.4 GHz band.

Notice that NGC~4261, IC~4296, and NGC~1316 have a lower $\mathrm{min}(t_{\rm{cool}}/t_{\rm ff})$ at a smaller radius than most of the other giant elliptical galaxies in the \citet{Lakhchaura2018} sample. Furthermore, the $t_{\rm{cool}}/t_{\rm ff}$ profiles in NGC~4261 and IC~4296, reach their minimum values in the central radial bin, raising the possibility that $\mathrm{min}(t_{\rm{cool}}/t_{\rm ff})$ is overestimated in these galaxies because of limited spatial resolution.  However, it is also possible for those $\mathrm{min}(t_{\rm{cool}}/t_{\rm ff})$ values to be slight underestimates.  In well-resolved galaxies that reach $\min(t_{\rm{cool}}/t_{\rm ff})$ outside the central radial bin, statistical fluctuations tend to cause the measurement of $\min(t_{\rm{cool}}/t_{\rm ff})$ to be biased low.  Figure~\ref{fig:tctff_profile} shows why the magnitude of that bias in the galaxies we are focusing on is likely to be small.  In all four galaxies, the second lowest value of $t_{\rm{cool}}/t_{\rm ff}$ is nearly identical to the minimum value, well within the 1-sigma statistical uncertainties.  Also, the $t_{\rm{cool}}/t_{\rm ff}$ profiles of those four galaxies are not constant in the 1--10 kpc range but only within the central $\sim 1$~kpc, where there are only a few radial bins, reducing the likelihood of an unrepresentative statistical fluctuation. Consequently, the fact that NGC~4261 and IC~4296 have unusually low $\mathrm{min}(t_{\rm{cool}}/t_{\rm ff})$ and greater radio power than most other galaxies in the sample, suggests that there may be a correlation between high radio power and $t_{\rm{cool}}/t_{\rm ff} < 10$ at small radii. In particular, the combination of low $\mathrm{min}(t_{\rm{cool}}/t_{\rm ff})$ and a power-law entropy slope that does not significantly flatten within the central kpc is a unique feature of NGC 4261 and IC 4296.  The other galaxies, in which the central entropy profile is flatter and $\mathrm{min}(t_{\rm{cool}}/t_{\rm ff})$ occurs at a larger radius, could be systems in which AGN feedback has recently elevated the entropy in the central kpc. 

\subsection{Comparison to Simulations}  
\begin{figure*}
\centering
  \includegraphics[clip,trim=2cm 0cm 2cm 1cm,width=0.7\textwidth]{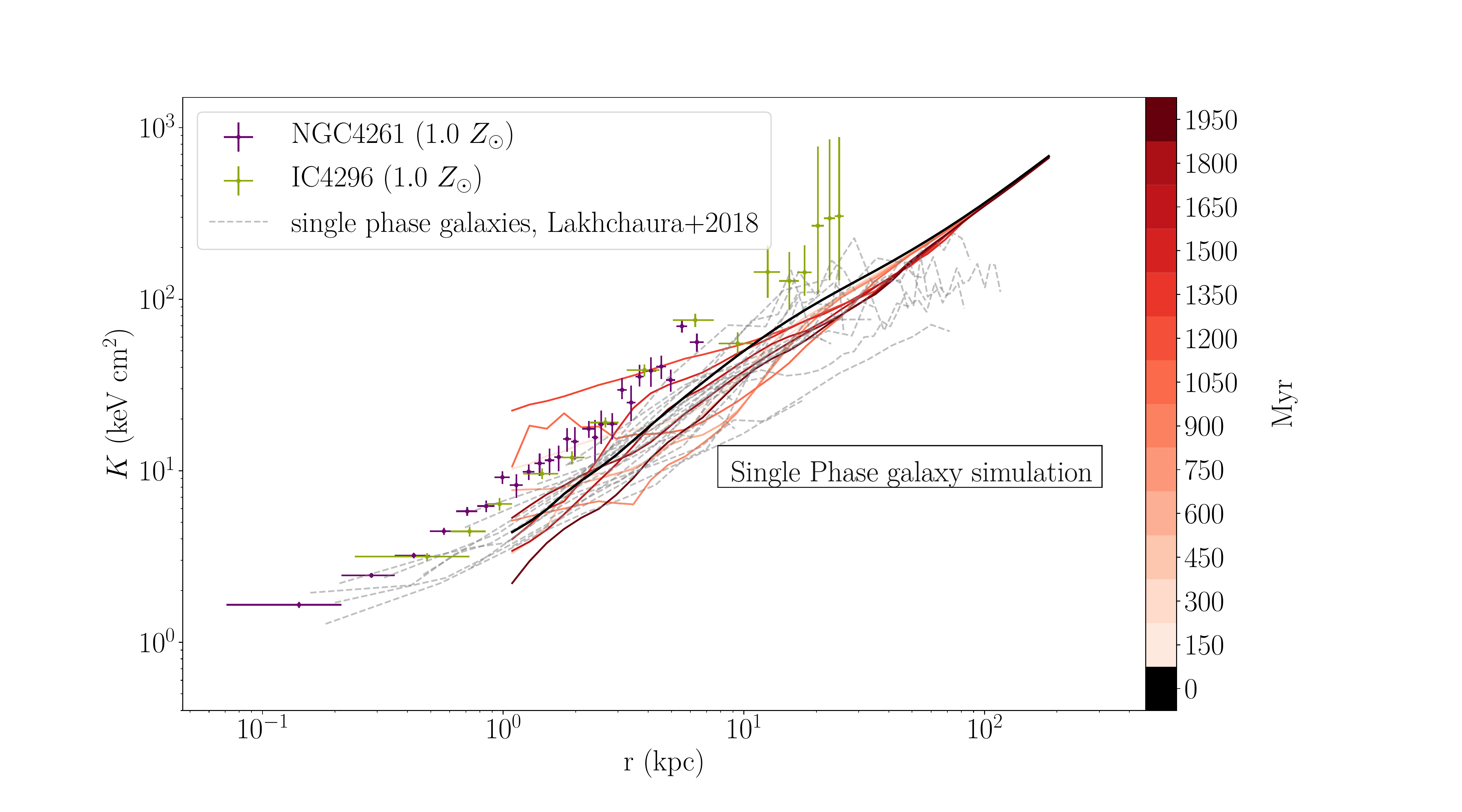}
  \includegraphics[clip,trim=2cm 0cm 2cm 1cm,width=0.7\textwidth]{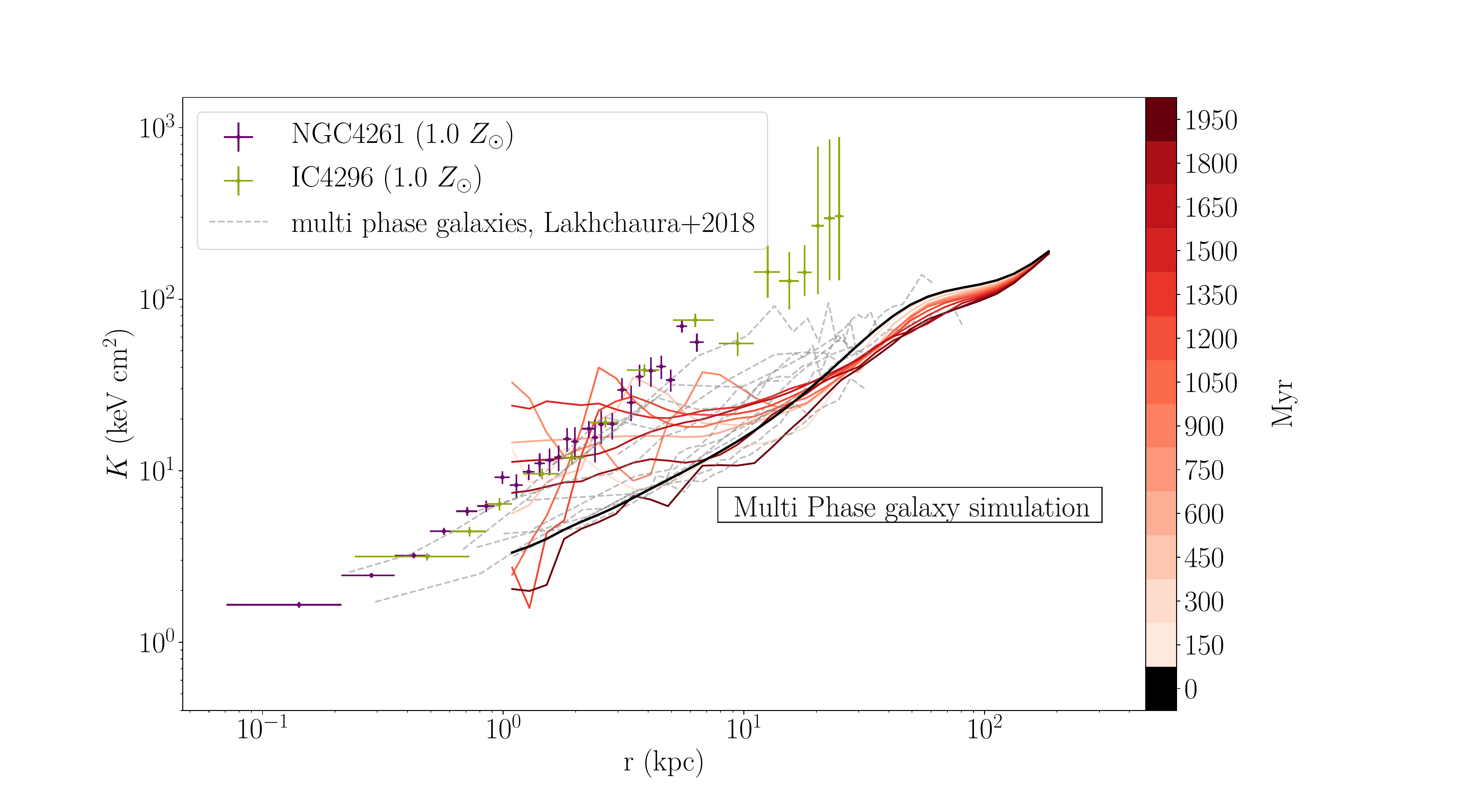}
  \caption{Entropy profiles of NGC~4261 and IC~4296 compared to simulations of somewhat lower mass giant elliptical galaxies with single phase gas (top) and multiphase gas (bottom) \citet{Wang2018} along with the single phase (top) and extended multiphase (bottom) galaxies from \citet{Lakhchaura2018}. Simulated profiles are shown at intervals 150 Myr, with earlier snapshots being shown in lighter red. The initial conditions are given by the black line and represent typical entropy profiles for single phase and multiphase galaxies, based on NGC 4472 and NGC 5044, respectively. Galaxies classified as ``N" (no extended multiphase gas) and ``E" (extended multiphase gas) from \citet{Lakhchaura2018} are included in grey with errors bars removed for clarity. The simulated galaxies have lower velocity dispersions than the observed galaxies, but the simulations were designed to represent the behavior over time of generic single and multiphase galaxies, rather than simulating a specific galaxy. The galaxies are referred to as MPG (multi phase galaxy) and SPG (single phase galaxy) instead of their names throughout \citet{Wang2018}. Note that the simulations do not resolve the gas profiles inside 1 kpc, and the flattening of the profiles in the center is likely a numerical effect because the resolution limit of the simulations can result in the simulated AGN affecting a larger region of the galaxy than real jets \citep{Wang2018}. Therefore, we expect the entropy slopes of the single phase simulations to be the same as our observations, though the normalization can differ. The measured entropy gradients are consistent with those seen in single phase gas simulations of radio sources in early-type galaxies.}
  \label{fig:Li_SPGMPG}
\end{figure*}

\citet{Voit2015b} showed that the presence of multiphase gas outside the central kpc of an early-type galaxy correlates with the slope of the entropy profile. Galaxies with an entropy slope of $K\propto r^{2/3}$ have multiphase gas present at $r>1~$kpc, while galaxies with an entropy slope of $K\propto r$ have only single phase gas beyond $r\sim1~$kpc. In order to better understand this relationship, we have compared our observed entropy profiles with the profiles of simulated galaxies from \citep{Wang2018}. Figure \ref{fig:Li_SPGMPG} shows a comparison between our data for NGC~4261 and IC~4296 and simulated elliptical galaxies with both single phase and multiphase gas as well as the entropy profiles for galaxies classified as having extended multiphase gas and no extended multiphase gas from \citet{Lakhchaura2018} . The initial conditions for the simulations are chosen to mimic X-ray observations of NGC~5044 (multiphase) and NGC~4472 (single phase), but the simulations were designed to represent generic multiphase and single phase galaxies. The simulations do not resolve the gas profiles at $<$ 1 kpc, meaning that our data have greater effective physical resolution than the simulations. However, we can still make comparisons in the $1-10$ kpc range. 

We begin by considering if the simulated galaxies are appropriate comparisons for our sample. From \citet{Makarov2014_vdisps}, the velocity dispersions of NGC~5044 and NGC~4472 are $224.9\pm 9.1\ \mathrm{km~s^{-1}}$ and $282\pm 2.9\ \mathrm{km~s^{-1}}$, respectively, while the velocity dispersion for IC~4296 and NGC~4261 are $327.4\pm 5.4\ \mathrm{km~s^{-1}}$ and $296.7\pm 4.3\ \mathrm{km~s^{-1}}$, respectively. \citet{Voit2015b} introduced the idea that there may be a correlation between the presence of multiphase gas, the velocity dispersion, and entropy profile slope in early-type galaxies. In contrast, \citet{Lakhchaura2018} found little correlation in entropy profile slope and the presence of multiphase gas when examining a larger sample. However, this is still an area of open study in both theory and observations of early-type galaxies. We do not expect the simulated profiles to match our data exactly because the velocity dispersions of the simulated galaxies are smaller than those of NGC 4261 and IC 4296.  However, we can still make useful comparisons between the overall behavior of the simulations and observations that account for how the entropy-profile slope correlates with velocity dispersion and the presence or absence of multiphase gas \citep{Voit2015b}.

For the multiphase and single phase initial conditions simulated by \citet{Wang2018}, the entropy profiles agree with the expectations of the simple physical models shown in \citet{Voit2015b}. The multiphase gas simulation has an entropy profile of $K(r) = 3.5 \, r_{\rm kpc}^{2/3} \, {\rm keV \, cm^2}$, which corresponds to the hypothesized precipitation limit at $t_c/t_{\rm ff}\approx10$.  The steeper entropy profile characteristic of single phase galaxies, $K(r) = 5 \, r_{\rm kpc} \,  {\rm keV \, cm^2}$, is consistent with models in which heating by Type Ia supernovae drives an outflow. The implication is that self-regulated AGN feedback can maintain the observed properties of both the single phase and multiphase galaxies, consistent with both idealized analytical models \citep{Voit2015b} and simulations \citep{Wang2018}.

When we compare the single phase simulation to NGC~4261 and IC~4296, the simulated galaxy does appear to maintain approximately the same entropy slope as our data. Furthermore, the comparison shows that the slopes of the single phase entropy profiles from \citet{Lakhchaura2018} are different from the slopes of the multiphase entropy profiles, and the multiphase data have similar slopes to the multiphase simulations. The comparison between NGC~4261 and IC~4296, the multiphase galaxy simulations, and the extended multiphase gas data, shows that our galaxies are better represented by single phase galaxies. The observed entropy profile slopes between 1-10 kpc are consistent with the entropy profile of a single phase galaxy, which has a steeper slope ($\alpha \sim ~ 1$). For some time steps, the central entropy profile of the simulated single phase galaxy flattens out, which could represent epochs when the black hole in the simulations is particularly active. However, it could also be a numerical effect because the resolution limit of the simulations can result in the simulated AGN affecting a larger region of the galaxy than real jets \citep{Wang2018}.

\subsection{Metallicity Analysis}\label{metallicity_analysis}
\begin{figure}
\centering
  \includegraphics[width=0.5\textwidth]{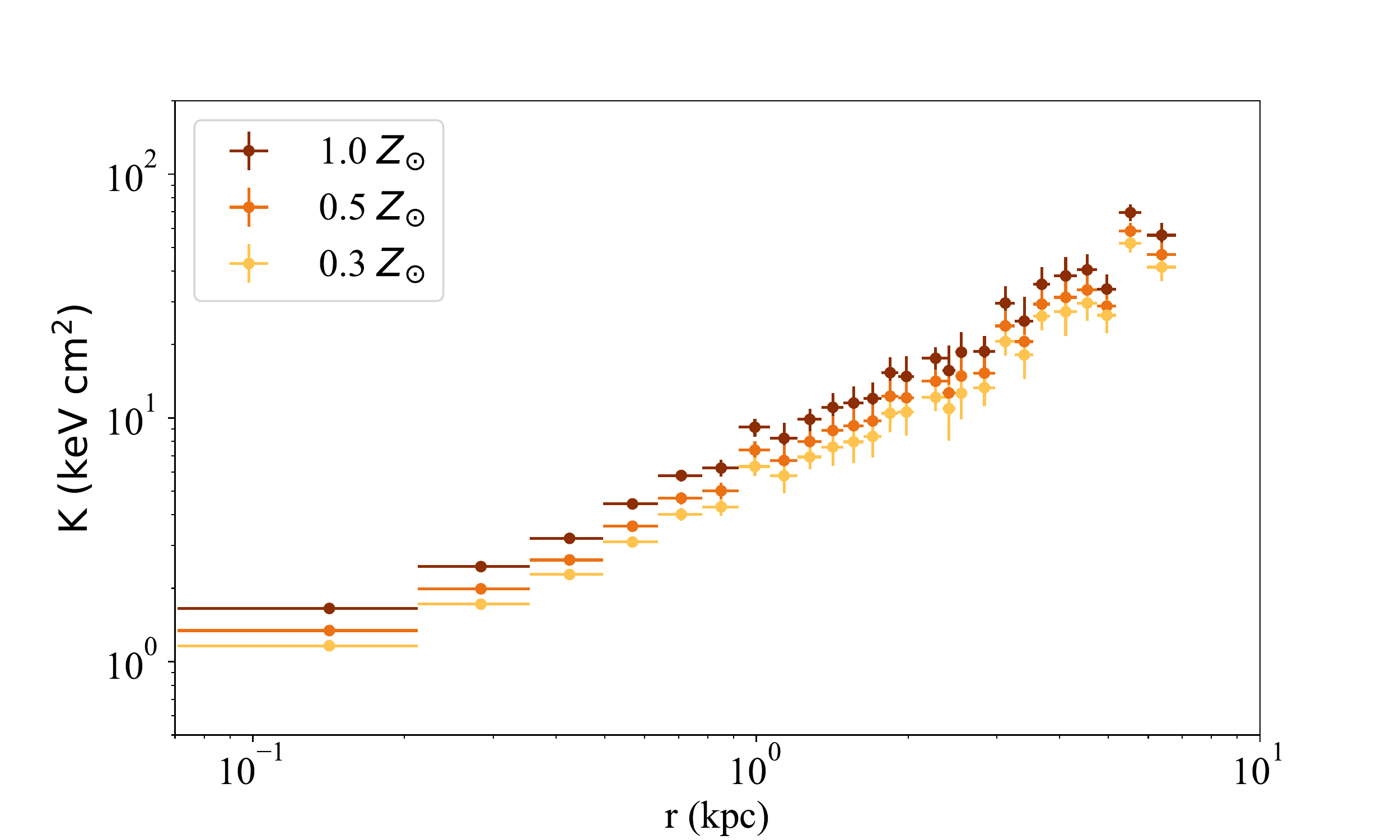}
  \caption{Comparison of the inferred entropy profiles for NGC~4261 for different assumed values of abundance. The points represent $1.0~Z_{\odot}$ (red), $0.5~Z_{\odot}$ (orange), and $0.3~Z_{\odot}$ (yellow). For increasing values of metallicity, the amplitude of the entropy profile increases. Therefore, if a galaxy has a different metallicity than we have assumed, and if we no longer assume that each galaxy has uniform metallicity, the slope of the entropy profile could change as well. However, we would still expect the profile to fall within the metallicity range illustrated in the figure.}
  \label{fig:Z_comparison_4261}
\end{figure}

When fitting entropy profiles, we assumed each galaxy had a constant metallicity of $1.0~Z_{\odot}$ across the profile. The hot gas abundances of early-type galaxies are difficult to obtain from low-resolution X-ray data, so we fixed the gas metallicities while fitting our observations. Here we quantify the sensitivity of our estimates of the X-ray densities and temperatures to the assumed metallicities. In Figure \ref{fig:Z_comparison_4261}, we show an example result of the impact of three different metallicity assumptions on the entropy profiles for NGC~4261: $0.3~\mathrm{Z}_{\odot}$, $0.5~\mathrm{Z}_{\odot}$, and $1.0~\mathrm{Z}_{\odot}$. This range of abundances spans those from various treatments of early-type galaxies in observations and simulations found in the literature \citep[][]{Werner2012,Wang2018,Li2014b}. The amplitude of the inferred entropy increases as assumed abundance increases, but the slope of the entropy profile shows little change. Therefore, abundance changes over the full range of expected gas abundances from $0.3-1.0$ solar would result in a change in amplitude of the inferred entropy profile on the order of 10-20\% (see Figure \ref{fig:Z_comparison_4261}). 

In comparing our results with entropy profiles from previous work, we find that different abundance assumptions indeed result in small differences in the inferred entropy profiles. However, when the identical assumptions for abundances are used, the entropy profile results for different authors are the same within the uncertainties. For example, the giant ellipticals examined by \citet{Werner2012}, the assumed abundances in the central kpc were $Z\sim0.5~Z_{\odot}$ and match our entropy profile results for NGC~4261 for a metallicity of $Z\sim0.5~Z_{\odot}$.  

The relation between abundance and electron density for an \texttt{apec} model for a narrow range of gas temperatures ($0.5-1.2$ keV) can be quantified approximately by 
\begin{equation}
\log{\left[\frac{n_e(Z)}{n_e(Z_{\odot})}\right]}=-m\log{(Z/Z_{\odot})}
    \label{eq:Z_ne}
\end{equation}
where $Z$ is metallicity assumed in the determination of $n_e$ and $m$ is the power-law slope. We would expect $m\sim 0.4$ based on the dependence of $\Lambda(T)$ on $Z$ in this temperature range (approximately $\Lambda \propto Z^{0.8}$) and form of the emission integral. 
To verify this estimate, we found the best-fit $n_e$ and $T_{\rm X}$ for the four galaxies with the best data quality, sampling a range of assumed metallicities. We determined that the best-fit temperature was insensitive to the metallicity assumption, while density and assumed metallicity were related as in Equation~\ref{eq:Z_ne} with:  
$m=0.43\pm0.04$ (NGC~1316), 
$m=0.43\pm0.04$ (NGC~4261), 
$m=0.39\pm0.11$ (NGC~4374), and 
$m=0.29\pm0.18$ (IC~4296), where uncertainties on $m$ are 1-sigma. These results are consistent with the expectations from X-ray plasma emission model for $kT\sim0.6-1$ keV.
The inferred electron density is therefore inversely related to the assumed abundance. 

Furthermore, if the abundance is actually lower in the center than we assume, we have underestimated the central density and overestimated the central entropy. So if a galaxy's gas is less metal-rich in the center than we have assumed, its central entropy profile could be slightly steeper than shown \citep[e.g.][]{Lakhchaura2019}. However, even if the central metallicity is lower than we assumed, the minima of the $t_{\rm{cool}}/t_{\rm ff}$ profiles are less than $t_{\rm{cool}}/t_{\rm ff} \,= \, 20$ for NGC~4261 and IC~4296.

\section{Conclusions}
Our analysis of the entropy profiles for a sample of nearby early-type galaxies with powerful radio sources shows that at least one other galaxy (IC 4296) is like NGC 4261 in having a powerful AGN and $t_{\rm{cool}}/t_{\rm ff} \sim$ 10 at $<$ 1 kpc. While the spatial resolution of the X-ray data for IC~4296 is not as good as for NGC~4261, both of their entropy profiles appear to be single power laws, and neither has extended multiphase gas greater than 2 kpc from their nuclei. To be certain of their similarity, we will need additional \textit{Chandra} observations of IC~4296 to match the data quality of NGC~4261. 

We produced deprojected temperature and density profiles for the hot gas surrounding seven additional early-type galaxies with powerful radio sources, but these observations lacked sufficient data quality to quantify the slope in the central $\sim0.5$~kpc. Unfortunately, these galaxies are likely not good candidates for further study at this time because the additional \textit{Chandra ACIS} observations needed to achieve comparable data quality to NGC~4261 are prohibitively long, given the degradation of {\em Chandra's} sensitivity to soft X-rays. We found that, in comparing independent analyses of entropy profiles in early-type galaxies, the treatment of abundance affects the amplitude of the entropy profile. Additionally, if the gas is not well-mixed, it may have a metallicity gradient, meaning that the slope of the profile could be affected as well. Finally, we compared IC~4296 and NGC~4261 to recent simulations \citep{Wang2018} and found that they are consistent with a single phase gas model galaxy. The simulations agree well with our observational results, providing positive evidence for their ability to robustly model the hot ambient medium in early-type galaxies.
In this work we were able to show excellent agreement between our observations, the theory of \citet{Voit2015b}, and the simulations of \citet{Wang2018}.  

\acknowledgments
{Support for this work was provided by the National Aeronautics and Space Administration through \textit{Chandra} Award Number SAO AR7-18008X issued by the \textit{Chandra} X-ray Center, which is operated by the Smithsonian Astrophysical Observatory for and on behalf of the National Aeronautics Space Administration under contract NAS8-03060.
The scientific results reported in this article are based on data obtained from the \textit{Chandra} Data Archive.
This research has made use of software provided by the \textit{Chandra} X-ray Center in the applications package CIAO. This research has made use of the NASA/IPAC Extragalactic Database (NED), which is operated by the Jet Propulsion Laboratory, California Institute of Technology, under contract with the National Aeronautics and Space Administration. The work of TC was carried out at the Jet Propulsion Laboratory, California Institute of Technology, under a contract with NASA.}

\software{Python \citep{van1995python}, SciPy \citep{scipy}, NumPy \citep{Numpy}, pandas  \citep{mckinney-proc-scipy-2010_pandas}, and matplotlib \citep{Hunter:2007_matplotlib}}






\bibliographystyle{yahapj}

\bibliography{ellipticalbib}

\end{document}